\def\fun#1#2{\lower3.6pt\vbox{\baselineskip0pt\lineskip.9pt
  \ialign{$\mathsurround=0pt#1\hfil##\hfil$\crcr#2\crcr\sim\crcr}}}
\relax
\documentstyle[12pt]{article}
\advance\textheight by \topskip
\begin{document}
\begin{titlepage}
\begin{flushright}
UH-IfA-94/68\\
December 1994
\end{flushright}
\vskip  .5 truecm
\begin{center}
{\Large\bf  STATISTICS OF COSMOLOGICAL}
\vskip  .5 truecm
{\Large\bf INHOMOGENEITIES}
\vskip 1.6cm
{\bf Lev Kofman}\\
\vskip .1cm
{Institute for Astronomy, University of Hawaii,
2680 Woodlawn Dr., Honolulu, HI 96822, USA}
\vskip .2cm
\end{center}
\vskip .2cm
\begin{quotation}
\centerline{\bf  ABSTRACT}
\vskip .2cm
This  contribution to the Proceedings is based on the talk given at the
Conference on Birth  of the Universe and Fundamental Physics,
 Rome, May 18-21, 1994. Some selected topics of the subject
are reviewed: Models of Primordial Fluctuations; Reconstruction
of the Cosmological Density Probability Distribution Function (PDF)
from Cumulants; PDFs in the Zel'dovich Approximation and
from Summarizing Perturbation Series; Fitting by the Log-normal
Distribution.
\vskip 1.3cm
\centerline{ In: {\it Proceedings of the Conference }  }
\centerline{ ``Birth of the Universe and Fundamental Physics''}
\centerline{ Rome, May 18-21, 1994; ed. F.Occhionero}
\end{quotation}
\end{titlepage}

\vfill
\eject

\section{Introduction}
\noindent
The canonical model for the formation of large-scale structure
is based on gravitational instability of small initial fluctuations
in the matter density $\delta$.
The primordial fluctuations are assumed to be a random field,
which is fully specified by the  joint probability density
functions (hereafter PDFs).
 These are assumed to originate
from quantum fluctuations of the inflaton scalar
field $\delta \phi_k e^{i k  x}$ that were stretched to
large comoving scales during the inflation phase \cite{1}.
For the simplest case of a single inflaton scalar field,
curvature (adiabatic) density perturbations
are generated by the inflaton fluctuations,
 $\delta \propto \delta \phi$,
which is a superposition of spatial waves $e^{i k x}$
with  random phases,
 realizes a random Gaussian field.

Theoretically,
there are scenarios based on non-Gaussian initial fluctuations.
Althogh these models currently do not appear to be our first choice,
they are interesting as alternatives.
 I will discuss these models in the next Section.
 The statistical nature of the
initial fluctuations
is therefore a basic distinguishing feature between  competing
theories.

The COBE DMR temperature anysotropy fluctuations sky map, in principle,
can  directly probe the statistics of the primordial fluctuations on large
 scales.
It is consistent with Gaussian fluctuations, but  its signal to noise
 level  so far is  sufficient only
to rule out strongly non Gaussian models \cite{2}.

In any case, at present the smoothed galaxy distribution is not a
Gaussian random field, due to the nonlinear dynamics of gravitational
 instability.  During linear evolution,
when all Fourier components $\delta_k$ evolve at the same rate, the  cosmic
PDFs do not
change form.  However,  even quasilinear evolution,
which makes phases correlated,  introduces strong
non-Gaussian features.

The  study of the PDFs of the large scale cosmic density and velocity fields,
 and
their moments  drew much attention   recently.
In Section 3, I will discuss recent analytic results on the derivation of the
 density and velocity PDFs  from gravitational dynamics.
There were suggestions     to design
the density PDF  phenomenologically \cite{3,4}. For instance,
 the log-normal form is a
surprisingly good  fit for
the densuty PDF in CDM model. Why it may be so, I will discuss
 in Section 4.

The discrete analogy of the one point PDF is the counts in cell
statistics. The  density PDF  is obtained by the
smoothing of the discrete galaxy field with a sharp filter.
 The observed density PDF manifests significant non-Gaussian
features in the  non-linear and even in the  quasi-linear regimes,
and can be well fitted by the log-normal distribution.
Assuming that galaxies trace mass, observed galaxy distribution
is  consisitent with Gaussian initial fluctuations on scales of
these surveys. However, the discriminatory power of the
observed PDFs is currently limited for several reasons. I will discuss these
 issues in the
Conclusion.

\section{Models of Primordial Fluctuations}

It has been shown that
in the framework of the
 the  inflation picture there is
still  room for non-Gaussian fluctuations.
In case of the curvature perturbations it requires
specific forms  of the inflaton potential or coupling with other fields,
plus  fine tuning of the parameters \cite{6}.
On the contrary, in case of the isocurvature perturbations
 generation of non Gaussian
fluctuations is rather typical \cite{6,7}.
Let another light scalar field $\chi$ be present at the inflation.
Long wavelengths
fluctuations $\delta \chi$ inevitably arise during inflation.
If later the $\chi$-particles become dominant and responsible for dark matter,
 then we come to the model with isocurvature perturbations.
The fluctuations $\delta \chi$ are Gaussian. However, they can generate
the isocurvature energy density perturbations
$\delta_{isocurv}=F(\delta \chi)$, which are non Gaussian for nonlinear
functions $F$ (local non Gaussian field \cite{5}).
An interesting example is the cosmic axion as $\chi$-field. The axion
energy density is proportional to the axion potential
 $V(\chi)=V_0 (1-\cos{\chi/\chi_0})$, therefore isocurvature fluctuations
in the model with axions as CDM are non Gaussian.
The power spectrum of isocurvature fluctuations does not depend on their
statistics. Unfortunately, the observed power spectrum ranging from the
horizon to the galaxy clustering scales leaves small room for the CDM scenario
with mixture of curvature and isocurvature fluctuations.

Another possibility is that the $\chi$-particles are underdominant but
their fluctuations are significant and play the role of the
 seeds for the structure formation \cite{7}. There are models
of the local non Gaussian isocurvature baryon fluctuations \cite{8}.
Non-Gaussian density fluctuations also arise in  scenarios,  where they
 originate from topological defects, such as cosmic strings \cite{9}
 or textures  \cite{13}, or late phase transition \cite{11}
or from non-gravitational cosmic explosions  \cite{12}.  Initial shape
$P(\delta)$ in scenarios with topological defects contains
 the long tail  in the overdense region, for instance,
 exponential for textures. It is interesting that the velocity PDF
in these model is normal, due to the central limit theorem \cite{13}.

\section{Reconstruction of  PDFs from  Cumulants}

First, let us recall basic terms which are adopted in the literature.
We use the cumulants of the distribution
$<\delta^p>_c$ rather than its
  moments  $<\delta^p>$.
The generating function of the cumulants  is
$C(\mu)=\sum_{p=2}^{\infty}<\delta^p>_c
{\mu^p/ p!}$,
where $\mu$ is an auxiliary parameter.
All cumulants of the normal distribution are vanishing, except the linear
 variance of the density fluctuations
 $\sigma$.
 It is convenient to use rescaled cumulants
$S_p(\sigma)=<\delta^p>_c/\sigma^{2(p-1)}$,
as a set of descriptive constants of a distribution.
The $S_p$ parameters would be constants in the hierarchical ansatz,
but in general they are functions of   $\sigma$.
The parameter $S_3$ multiplied by $\sigma$ is the skewness,
and $S_4$ multiplied by $\sigma^2$ is the kurtosis.
The advantage to use these parameters is that they are  final numbers
 $S_p(0)$
for an arbitrarily small   $\sigma$. The parameters  $S_p(0)$
are the fingerprints of the particular dynamical model.

When the whole series of the cumulants is known it is possible
to reconstruct the density PDF from the
generating function of the cumulants
\begin{equation}
P(\rho)= {1 \over 2\pi i}\int_{- i\infty}^{+ i\infty}
d \mu
\exp\left[-{C(\mu)}-{(\rho-1)\mu}\right].
\label{recon}
\end{equation}
Here and further $\rho$ is the density in units of mean density.
 In practice we may have only a few lower cumulants,
 such as the skewness and the kurtosis. Theoretically, they can be calculated
in the perturbation theory.
In the case of the weakly non-linear dynamics, when a slight departure from the
initial
Gaussian distribution is expected,
 one can obtain the general decomposition
series around the Gaussian PDF
induced by the first non-zero cumulants
-- the Edgeworth expansion.
In cosmological context this was implicated in \cite{14,15,16}.

The  Edgeworth expansion for the cosmic density PDF
can be obtained from the reconstruction formula (\ref{recon}), keeping a few
lowest terms in the generating function $C(\mu)$. The result is
\begin{equation}
P(\delta)={1 \over (2\pi\sigma^2)^{1/2}}  e^{-{{\delta^2} \over {2\sigma^2}}}
\biggl[1 + \sigma {S_3 \over 6} H_3\left(\nu\right)
+\sigma^2\biggl( {S_4 \over 24} H_4\left(\nu\right)
+{S_3^2\over72}H_6\left(\nu\right)
 \biggr)
+ ... \biggr]
\label{edge}
\end{equation}
where $H_n(\nu)$ are the Hermite polynomials, $\nu=\delta/\sigma$.
The actual forms of the parameters $S_p=S_p(\sigma)$ which depend
 on the particular dynamics,  affect the  expansion (\ref{edge})
with respect to $\sigma$.
The usual  measurements of  the lowest cumulants are significantly
affected by  the high density tail of the PDF, i.e. the rare events.
It is interesting, that  the lowest
cumulants alone are responsible for the shift of the peak of
$P(\delta)$. Therefore the measurement of the shape of the PDF maximum,
which statistically is more robust,
can provide an alternative method of evaluation of the lowest cumulants.
The
Edgeworth expansion
 fails to reproduce  $ P(\delta)$ when
$|\delta| \geq 0.5$ where spurious wiggles appear.
In practice it is useless for  $\sigma \geq 0.3$.
The
Edgeworth expansion also can be used for modelling slightly non Gaussian
initial fluctuations.

\section{PDFs in the Zel'dovich Approximation}

Physical meaning of the density PDf is the fraction of volume with
a given level of density.
The former shape of the initial density PDF is broken as non-linearity
develops, since matter is evacuated from the underdense regions, which
occupy the largest fraction of volume, meanwhile the overdence regions
 collapse.
In the formal expression (\ref{edge}) this corresponds to the
 positive sign of the skewness, increasing with time.
This feature of the density field  evolution is clearly manifested
in the Zel'dovich approximation
\begin{equation}
\rho   =  {{\rho_0}
 \over {\vert(1-D\lambda_{01})(1-D\lambda_{02})(1-D\lambda_{03})\vert}},
\label{zeld}
\end{equation}
where  $\rho_0$ is  the initial local density,
 growing mode of adiabatic perturbations is $D(t)$,
and  $\lambda_{0i}$ are  the eigenvalues of
the Lagrangian  deformation tensor.
This formula does not assume any particular initial statistics, whether
Gaussian or not.
The gravitational clustering at  sufficiently large  scales
  can be
considered  in the quasilinear theory  in a single stream regime
ignoring small scale details.
The Zel'dovich approximation (\ref{zeld})  can be  applied  for the
filtered  initial gravitational potential.
This approach sometimes is reffered to as the truncated Zel'dovich
 approximation.

In the truncated Zel'dovich approximation
 the statistics of the evolved density field can then entirely be
obtained from the statistics of the initial local density $\rho_0$
and the initial eigenvalues $\lambda_0$-s. For adiabatical
perturbations  $\rho_0$ is the background mean density.
Let  the initial joint PDF of all involved cosmic fields, including
velocity and gravitational potential, be
$W_0(\rho_0,\lambda_{01},\lambda_{02},\lambda_{03},
 \vec u_0, \Phi_0)$.

The density PDF can be obtained in the general case of arbitrary
 initial statistics  simply  as an integral
over all involved variables except density \cite{15,16}
\begin{equation}
P(\rho)=\int  \delta \bigl[ \rho
\vert (1-D\lambda_{01})(1-D\lambda_{02})(1-D\lambda_{03})\vert-  1\bigr]
W_0  d\lambda_{01} d\lambda_{02} d\lambda_{03}
 d^3u_0 d\Phi_0,
\label{nogauss}
\end{equation}
For the  Gaussian initial conditions we can omit
$\vec u_0$ and $\Phi_0$ in the initial joint PDF and write it  as
$ W_0( \rho_0, \lambda_{01}, \lambda_{02}, \lambda_{03} )=
 \delta (\rho_0-1)
 M_0( \lambda_{01}, \lambda_{02}, \lambda_{03})$.
The second factor is the well known joint distribution function
of the eigenvalues of the initial deformation tensor for
an initial Gaussian displacement field \cite{17}.
Substituting this form of $W_0$ into the integral (\ref{nogauss}),
after some tedious algebra, we can reduce it to the
 one-dimensional  integral \cite{18,19}:
\begin{equation}
P(\rho)=
{N \over \rho^3}
\int_{3 ({\bar \rho \over \rho })^{1/3}} ^{\infty}  d s\
e^{-{(s-3)^2 / 2 \sigma^2}}
\left( 1+ e^{-{6s/ \sigma^2}} \right)
\ \left( e^{-{\beta_1^2 / 2\sigma^2}}
   +e^{-{\beta_2^2 / 2\sigma^2}}
   -e^{-{\beta_3^2 / 2\sigma^2}}  \right)
\label{zel}
\end{equation}
where
$\beta_n (s) \equiv s \cdot \,5^{1/2} \left( {1\over2}
+\cos\left[{2\over3}(n-1)\pi
+{1\over3} \arccos \left({{54{ \bar \rho}^3} \over \rho s^3}
-1 \right)\right]\right)$, and
prefactor is $ N={{9 \cdot 5^{3/2}} \over {4\pi   \sigma^4}}$.
In the limit of small $\sigma$ the expression (\ref{zel}) transfers into the
form (\ref{edge}).
The density PDF calculated numerically from formula (\ref{zel})
is plotted in Fig. 1. Without final smoothing, this PDF does not depend
on the power spectrum $n$. The quality of approximation of actual PDF by
the formula (\ref{zel}) is increasing as $n$ is decreasing $n \to  -3$.
 The $n$-dependence can be taken into
account in the limit of small $\sigma$ \cite{20}.

For the slightly non Gaussian initial fluctuations, one can expand the
joint distribution $W_0(\rho_0,\lambda_{01},\lambda_{02},\lambda_{03},
 \vec u_0, \Phi_0)$ around its Gaussian form, discussed above.
For this purpose it is convenient to return to six components of the
deformation tensor, and apply the generalized Edgeworth expansion
for several variables. From (\ref{nogauss}) one can obtain the
density PDF evolving in time from slightly non Gaussian fluctuations
in form of series around distribution (\ref{zel}). One of the  lesson
from this exercise is that, in principle, the final density statistics
 depends on
the statistics of all fields involved in
the joint distribution $W_0(\rho_0,\lambda_{01},\lambda_{02},\lambda_{03},
 \vec u_0, \Phi_0)$.\\

\section{ PDFs from Summarizing Perturbation Series}

Using the perturbation theory  for Gaussian initial conditions, it is
 possible to derive the cumulants of the density PDF
in  the small $\sigma$ limit.
 The basic assumption
here is that  the gravitational clustering
at sufficiently large scales can be considered in the single stream regime.
Derivation of the lowest cumulants in the quasi-linear dynamics
in the single stream regime was extensively elaborated \cite{20,21,22,23,24}.
Bernardeau \cite{25,26}
found an elegant method to derive the closed form for the
generating function of the cumulants $C(\mu)$ in the limit of small
$\sigma$, which allows to obtain all parameters $S_p(0)$ and
reconstruct the density PDF in this regime.

The chain of
equations for  density perturbation $\delta^{(p)}$ in
 each order $p$ cannot be resolved.
However, one can define the   connected  averages:
 $\nu_p \equiv <\delta^{(p)} (\delta^{(1)})^p >_c/\sigma^{2p}$, and
 construct their generating function
$G(\tau)=\sum_{p=1}^{\infty}(-\tau)^p {\nu_p / p!}$,
where $\tau$ is an auxiliary variable.
 It is remarkable that there is  a single equation for $G(\tau)$
which corresponds to
the density contrast of the spherical collapsing linear overdensity $\tau$
 \cite{29}.
An approximated analytical solution for this function is
\begin{equation}
G(\tau) = {1\over ( 1+\tau/\alpha)^{\alpha}}-1.
\label{solution}
\end{equation}
For the three dimensional
single stream cosmological gravitational instability $\alpha \approx 1.5$
\cite{26}.

This method and results are rather general
in the limit of small $\sigma$. For instance \cite{16},
for the Zel'dovich dynamics in the one dimensional case, where the Zel'dovich
 ansatz is an exact solution, the formula (\ref{solution}) is valid with
$\alpha =1$. For the Zel'dovich approximation in the two and three dimensional
cases, $\alpha = N$, where $N$ is the space dimension.
In the formal limit $N \gg 1$ we get $G(\tau)=\exp(-\tau) -1$,
which coincides with  that of the log-normal distribution.

The transition from $G(\tau)$ to the generating function of the cumulants
$C(\mu)$ is given by
 the Legendre transform
with the variable of the transform equal to unity afterwards \cite{32,33}.
Substituing the solution (\ref{solution}) into (\ref{recon}), one
 reconstructs the density PDF.
Without final smoothing,
this PDF depends on $\sigma$ only and not on the power spectrum index $n$.
However, for a practical purpose it is necessary to filter
the evolved  density field. After filtering, density cumulants and
PDF depend   on the shape of the power spectrum \cite{22,24}.
Bernardeau demonstrated \cite{27}, that
the effect of filtering
reduces to the a simple transformation of the generating function
$G(\tau) \to G^{f}(\tau)$.
 The resulting shape of the density
PDF, based on the filtered generating function
 $G^{f}(\tau)$, is presented in Fig. 1, 2.

\section{Fitting By The Log-Normal Distribution}

As it was noted a long time ago by Hubble \cite{30}, the galaxy count
distribution in angular cells on the sky might be well described
by a log-normal distribution.
The recent study of the spatial distribution of galaxy count
are in good agreement with the log-normal fit
\cite{31,32,19}
The log-normal density distribution reads
\begin{equation}
P(\rho)=
{1\over\left(2\pi\sigma_l^2\right)^{1/2}}
\exp\left[-{\left(\ln{\rho}+\sigma_l^2/2\right)^2\over
2\sigma_l^2}\right]{1 \over \rho},\ \ \sigma_l^2=\ln
(1+\sigma^2).
\label{lognor}
\end{equation}
It was  found \cite{23,20} that the log-normal distribution is an excellent
approximation to the density PDF from N-body CDM simulation for
 the tested values of $0.3 < \sigma <1.5$  in the
tested range of $\rho \le 10$, see Fig. 1.
Such a remarkable fitting inspires the thought that there might be a
 dynamical reason to manifest the log-normal features of the
density PDF. In \cite{3} the log-normal
mapping of the linear density field was suggested
 to describe its non-linear evolution.
This log-normal model is universal for any spectral index $n$.
Unfortunately the log-normal mapping does not work \cite{33}.
At Fig. 2 the log-normal distribution is compared with the analytically derived
density PDF \cite{34}.
 The log-normal shape fits an actual PDF not for arbitrary $n$,
but for $n \approx -1$.
It is easy to understand, considering
 the cumulants of the log-normal distribution. For instance, we have
$S_3^{log}(\sigma)=3+\sigma^2$
for an arbitrary $\sigma$.

The actual  parameter is $S_3(0)={34 \over 7}-(n+3)$. The parameter
 of the log-normal distribution $S_3^{log}(\sigma)$ fits an actual parameter
not for arbitrary $n$ and $\sigma$, but along  the curve
 $n = -1.14 - \sigma^2$. The $\sigma$-correction to the $S_3(0)$ changes
slightly
the numerical prefactor of $\sigma$. The parameter $S_4^{log}(\sigma)$ fits
an actual parameter  $S_4(\sigma)$ approximately along the same
curve. Thus,
the log-normal distribution fits well in the particular region of
the parameter space $(n,\sigma)$ around the ``log-normal'' curve $n(\sigma)$,
but worsening outside of this region. The actual density PDF can
be further approximated by the Edgeworth expansion around the log-normal form
\cite{35}.
 By chance, the popular
CDM model at moderate $\sigma$ crosses  this region.

Equipped with this method, one can understand why and where the other
 fits such as ``thermodinamic'' \cite{4,36} or the negative
binomial distribution \cite{32,37} are applicable.

\section{Discussion}

In general case, $S_p$-parameters as functions of the linear density
variance can be presented  as series
\begin{equation}
S_p(\sigma, n)=S_p(0, n)+T_p(n)\sigma^2 + ...
\label{sigma}
\end{equation}
There is derivation of all coefficients $S_p(0, n)$  \cite{25}, but a little
is known about $T_p(n)$, which describe a lowest $\sigma$-correction.
Apparently,  $T_p(n)$ depends on $n$.
Some N-body simulations indicate the tendency  $S_p(\sigma, n)$
 to increase with $\sigma$ faster
as $n$ decreases \cite{38}.
Parameters $S_p$
 slowly increase with $\sigma$ at quasilinear regime,  $T_p(n) \ll S_p(0)$,
 at least for
  $ -1 \leq n \leq 0$ \cite{14}
and for CDM model \cite {16}.
A possible explanation is that the spherical collapse
is the dominanted form for these spectra, contrary to
 the pancakes for $n \to -3$.
Another choice of the set of descriptive parameters is
 $<\delta^p>/\sigma_{nl}^{2(p-1)}$, where nonlinear  variance is
$\sigma_{nl}(\sigma)$. It was shown in \cite{41} numerically for CDM model
 that these combinations remain constants equal to $S_p(0)$ for a wide
range of $\sigma_{nl}$.
 Therefore Bernardeau's method, formally
applicable for small $\sigma$,
 can be extrapolated
across the whole quasilinear regime, which makes it very useful
for practical application.
 The $\sigma$-dependence in $S_p$
 can be more significant for $n \to -3$. In this case the accuracy of
 PDF based on the Zel'dovich approximation, is increasing.
Another related question is the hierarchical ansatz in the highly nonlinear
regime. It assumes that $S_p(\sigma)$ are saturated
 as  $\sigma \gg 1$. This is still an unanswered question, however,
there are interesting results in this regime \cite{39,40}.

In practice,
galaxy count in cell distributions, and consequently,
 the density PDFs have been measured in various catalogs in
many works, for example see  \cite{31,37,42}
for the CfA survey,   \cite{32,19} the IRAS surveys,   \cite{42} for
SSRS survey,  in {43} for clusters of galaxies, in \cite{44} for
$S_p$ parameters from the APM survey,
 and references therein.
 These data are compatible with Gaussian initial fluctuations, and can be
 fitted  by the log-normal distribution, see, e.g., Fig. 3.
Thus, some ``log-normal'' features of the observed density PDF would mean
that the realistic cosmological model has the $n(\sigma)$ dependence
close to the CDM-type models.
Unfortunately, the error bars of the observed density PDF
  increase at large $\rho \sim 2-3$, and significantly
 affected by the depth of the sample \cite{32,23}. Therefore currently
 one can rule out only
strongly non Gaussian models.
Another potential problem is galaxy bias, which affects the observed deviation
from Gaussianity \cite{45}. The linear bias at filtering
scales would  preserve the reported analytic results. However,
 the departure from
 this simple rule is expected, as predicted by numerical
 simulations \cite{46}.

The density PDF deviates from a normal distribution very rapidly.
On the contrary, the velocity PDF departures from its initial distribution
very slowly \cite{19,47}. The observed velosity PDF is shown at Fig. 3.

\section{Acknowledgements}

I  thank Prof. Franco Occhionero and the Organizing Committee of
the Conference for warm hospitality  support in Rome.

\vfill
\eject

{ \bf Fig. 1}  The density PDF obtained by
different methods for CDM initial conditions \cite{20}. The
points are measured in a numerical simulation at the
time  corresponding to the present epoch,
  at two different smoothing radii $R_0=5h^{-1}$Mpc
(triangles) and $R_0=15h^{-1}$Mpc
(circles). The rms density fluctuation are then respectively
 $\sigma=1.52$
and $\sigma=0.47$. The error bars have been obtained by
dividing the sample into eight subsamples. The solid line is the prediction
 of the Bernardeau's method when the smoothing effects are taken into account.
The dashed line is the prediction (\ref{zel}) from the ZA and
the long dashed line is the lognormal distribution
 (\ref{lognor}).\\

{ \bf Fig. 2} The dynamically motivated density PDF for
 different $n$,   $\sigma=0.5$, versus log-normal distribution \cite{34}.
The solid line for $n=+1, 0, -1$ is the density
PDF obtained by the Bernardeau method with the final smoothing.
The solid line for $n=-2$ is the PDF in the Zel'dovich approximation.
The  dashed line is the  log-normal distribution.\\

{ \bf Fig. 3}   PDFs for IRAS 1.9Jy density and velocity fields
 in a sphere of radius $80 h^{-1} Mpc$,
after Gaussian filtering with  $R_s=6 h^{-1} Mpc$.
Dashed and long dashed curves are the Gaussian and lognormal
 distributions with the same
$\sigma$.
Also shown the errors associated with the limited volume sampled
\cite{19}.

\end{document}